\documentclass{ws-ijmssc}

\begin{document}

 \newcommand{\be}{\begin{equation}}
\newcommand{\ee}{\end{equation}}
\newcommand{\bq}{\begin{eqnarray}}
\newcommand{\eq}{\end{eqnarray}}
\newcommand{\bsq}{\begin{subequations}}
\newcommand{\esq}{\end{subequations}}
\newcommand{\bc}{\begin{center}}
\newcommand{\ec}{\end{center}}
\markboth{T. A. PEREIRA, J. MENEZES AND L. LOSANO}{INTERFACE NETWORKS IN MODELS OF COMPETING SPECIES}

\catchline{}{}{}{}{}

\title{INTERFACE NETWORKS IN MODELS OF COMPETING SPECIES}

\author{T. A. PEREIRA}

\address{Institute for Biodiversity and Ecosystem Dynamics, University of Amsterdam, Science Park 904, 1098 XH Amsterdam, The Netherlands\\
Departamento de F\'isica Te\'orica e Experimental, Universidade Federal do Rio Grande do Norte, 59078-970, RN, Brazil.\\
tiberio@dfte.ufrn.br}

\author{J. MENEZES\footnote{Corresponding author.}}

\address{Institute for Biodiversity and Ecosystem Dynamics, University of Amsterdam, Science Park 904, 1098 XH Amsterdam, The Netherlands\\
Escola de Ci\^encias e Tecnologia, Universidade Federal do Rio Grande do Norte\\
Caixa Postal 1524, 59072-970, Natal, RN, Brazil\\
jmenezes@ect.ufrn.br}

\author{L. LOSANO}

\address{Departamento de F\'{\i}sica, Universidade Federal da Para\'{\i}ba 58051-970 Jo\~ao Pessoa, PB, Brazil, 
Caixa Postal 1524, 59072-970, Natal, RN, Brazil\\
losano@fisica.ufpb.br}

\maketitle

\begin{history}
\received{(Day Month Year)}
\accepted{(Day Month Year)}
\end{history}

\begin{abstract}
We study a subclass of the May-Leonard stochastic model with an arbitrary, even number of species, leading to the arising of two competing partnerships where individuals are indistinguishable.
By carrying out a series of accurate numerical stochastic simulations, we show that alliances compete each other forming spatial domains bounded by interfaces of empty sites.
We solve numerically the mean field equations associated to the stochastic model in one and two spatial dimensions. We demonstrate that the stationary interface profile presents topological properties which are related to the asymptotic spatial distribution of species of enemy alliances far away from the interface core. Finally, we introduce a theoretical approach to model the formation of stable interfaces using spontaneous breaking of a discrete symmetry. We show that all the results provided by the soliton topological model, presented here for the very first time, are in agreement with the stochastic simulations and may be used as a tool for understanding the complex biodiversity in Nature.  
\end{abstract}


\section{Introduction}

\label{intro}
It is well known that the interactions among species are responsible for the large variety of biodiversity observed in Nature \cite{sole2006selforganization,nowak06evolutionaryDynamicsBOOK}. For example, spatial patterns reveal that species may form alliances in order to promote their coexistence \cite{doi:10.1080/17429145.2011.556262,Szabó200797}.
Early studies in mathematical biology have shown that population dynamics in the well-mixed scenario can be understood through a set of differential equations \cite{doi:10.1021/ja01453a010,Volterra}. When it comes to space, however, the mathematical analysis of the dispersal of species becomes more difficult. For this reason, numerical stochastic simulations have played an important role on the investigation of spatial population and ecosystem \mbox{dynamics \cite{May-Leonard,Kerr2002,Reichenbach2007,PhysRevLett.77.2125,PhysRevLett.99.238105,Leisner2012,Cheng2014,Daly2015189,PhysRevE.89.012721,2014JPhA...47p5001M,1367-2630-17-11-113033,Weber20140172,PhysRevE.91.033009,PhysRevE.91.052135,PhysRevE.92.022820,Roman201610,PhysRevE.84.021912}}.  
As an example, stochastic simulations of the standard rock-paper-scissors game (three strategies) describe the population dynamics
of coral reef invertebrates \cite{coral} and lizards in the inner Coast Range of California \cite{lizards}.  The numerical results are also in agreement with experimental tests using microbial laboratory cultures of three strains of colicinogenic \textit{Escherichia coli}  \cite{bacteria}.
In the case of two competing species, stochastic methods can be used to model interactions between butterflies \cite{but1,but2}. Here, the spatial patterns show the formation of regions occupied by groups of distinct species, bound by interfaces of empty spaces, whose \mbox{dynamics} is similar to \mbox{topological} defect networks in cosmology and condensed matter \cite{Boerlijst199117,PhysRevE.87.032148,0305-4470-38-30-005,PhysRevE.64.042902,Lutz2013286,PhysRevE.86.031119,PhysRevE.86.031119,PhysRevE.86.036112,PhysRevE.89.042710,Avelino2014393}. Specifically, the interface network length scale  $L$ grows as $L \propto t^{1/2}$, which characterises a scaling regime \cite{PhysRevLett.98.145701}.

Formation and stability of alliances in Biological systems are well studied in
experimental settings. For example, plants and mites form partnerships, which allow
them to coexist with other species\cite{doi:10.1080/17429145.2011.556262}. On the other hand, in the theoretical scenario, various
models have been proposed to explain the interaction between the species and the
spatial distribution of individuals\cite{PhysRevE.64.042902,PhysRevE.89.042710}.
Our purpose is to provide a general model, where members
of partnerships have the same role of protecting the partners and expand the territory
occupied by the alliance. This may be of interest to biologists, that would apply the
spatial modeling to describe systems of alliances where different species work together\cite{but2}. 

In this paper, we focus on the topological aspects of interface networks in systems with two partnerships. Although some authors have presented some mechanisms of formation of alliances \cite{PhysRevE.64.042902,Lutz2013286,PhysRevE.86.031119,Menezes2017}, there is room to investigate partnerships where individuals of different species are indistinguishable. We present the results of the stochastic numerical simulations and highlight the novelty of the model in contrast with the models of partnerships in literature. Besides, we introduce \mbox{differential} equations and run deterministic simulations to describe the dynamics of the spatial distribution of the species. 
Because of the strong similarity with topological defects, we subsequently introduce a soliton topological model to describe the stationary profile of the interfaces. We concentrate on spatial patterns formed through a phase transition driven by a spontaneous breaking of the discrete symmetry \mbox{associated} to two alliances. This new formalism may allow a direct comparison of the population dynamics with topological defect networks extensively studied in other scenarios\cite{Stavans1989,Flyvbjerg1993,Monnereau1998,Weaire2000,Avelino2008,Avelino2010,PhysRevLett.101.087204}.

In the next section, we introduce the spatial stochastic model with an arbitrary, that leads to the formation of two partnerships of equal \mbox{individuals}. In Sec. III, the mean field equations are introduced, and the dynamics of the interface networks is studied. In Sec. IV, the soliton topological model is presented, and the stationary profile of the interface is found. In Section V we discuss the results by comparing the theoretical soliton topological model with the numerical implementation of the mean field equations. Finally, our main conclusions are presented in Sec. VI. 

\begin{figure}[t]
	\centering
	\includegraphics*[width=3cm]{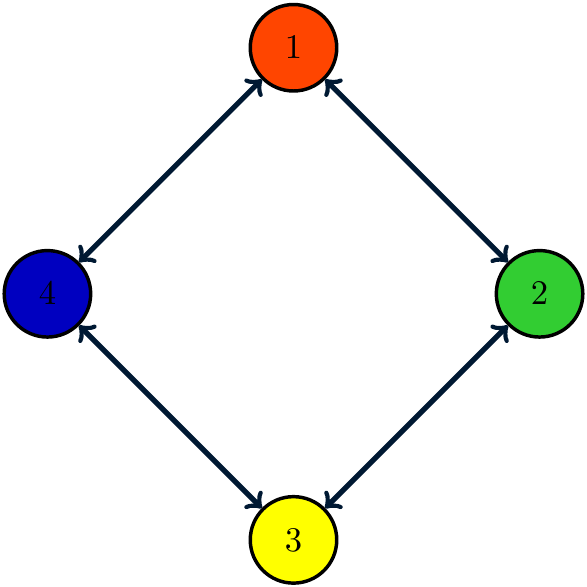}
	\includegraphics*[width=3cm]{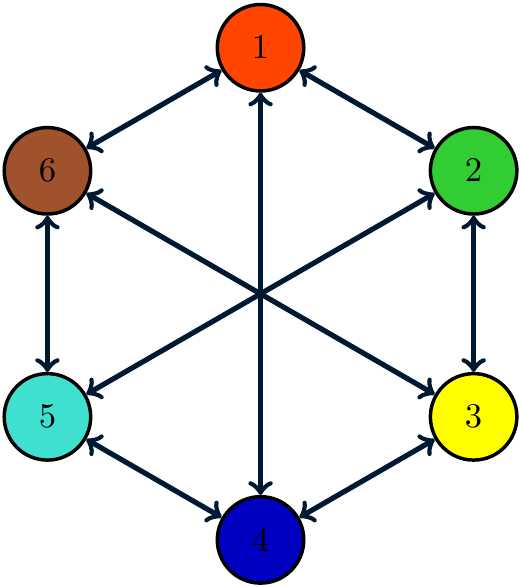}
	\includegraphics*[width=3cm]{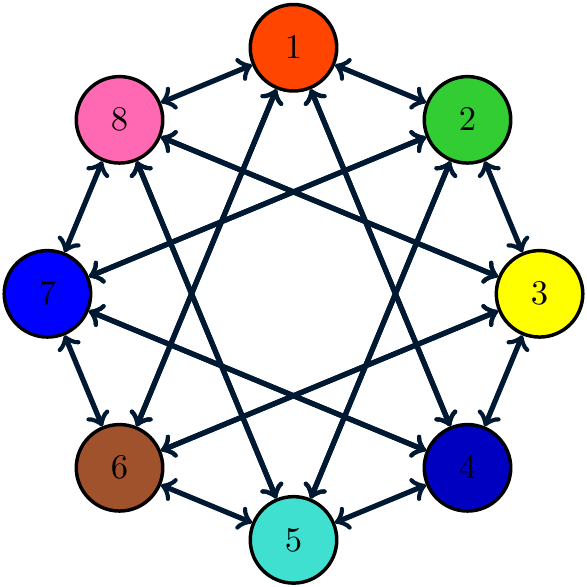}
	\includegraphics*[width=3cm]{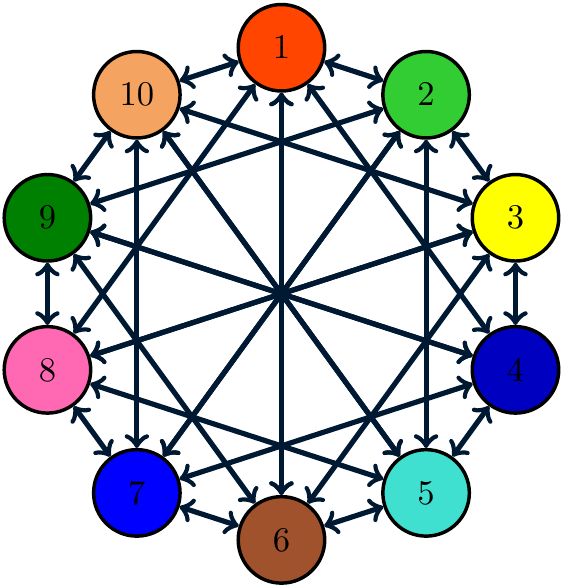}
\caption{Illustration of the competition schemes for the models with $4$, $6$, $8$, and $10$ species (from left to right, respectively). }
 \label{fig1}
\end{figure}
\section{The Stochastic Model}
We consider a class of models with an even number $N$ of species, where individuals of species $i$ compete with individuals of $N/2$ other species. This means that we aim
to investigate a sub-family of the general spatial stochastic May-Leonard models \cite{May-Leonard}, where species segregate into two alliances of $N/2$ species.
In our stochastic model, there are three types of \mbox{interactions}: mobility
($ i\ \odot \to \odot\ i\,$), reproduction ($ i\ \otimes \to ii\,$) and selection (competition or predation) ($ i\ j \to i\ \otimes\,$), with $j=i+k$ where $k$ is an odd number $k<N$. Note that while ($\otimes$) means an empty site, $\odot$ may be a vacancy or an individual of any species. Competition creates empty spaces, i.e., whenever one individual dies, the grid point is left empty. Subsequently, this vacancy will be occupied by a new individual created by reproduction of any species. 

From left to right, Fig.~\ref{fig1} illustrates the competition rule for $N=4$, $N=6$, $N=8$, and $N=10$. Although each species is labelled by $i$ (or $j$), the cyclic identification $i=i+l\,N$, where $l$ is an integer is assumed. Moreover, all species have the same diffusion, reproduction and competition rates. This implies that, except for the labeling of the different species, schemes in Fig.~\ref{fig1} are invariant under a rotation of $2\,\pi/N$, leading to a $Z_N$ symmetry.

We performed a large number of stochastic numerical simulations taking two-dimensional networks with $\mathcal{N}$ grid points and periodic boundary conditions. Initially, the number density, $n_i=I_i/\mathcal{N}$ is assumed to be the same for all species whereas no empty site is present $n_E=I_E/\mathcal{N}=0$  ($I_i$ and $I_E$ are the total amounts of individuals of species $i$ and empty sites, respectively).  At each time step, a random individual is sorted to interact with one of its neighbors. After $N$ interactions take place, one generation (our time unit $\Delta t=1$) is computed. 

All snapshots of the network simulation presented throughout this paper, obtained by assuming $m=0.25$, $p=0.50$ and $r=0.25$, were found to be adequate for visualization purposes. Nonetheless,  we verified that many other choices of the parameters would provide similar qualitative results. Furthermore, we will show later in this paper that different parameters lead to different quantitative results -
the interface thickness and height are functions of $m$, $p$, and $r$.

\subsection{Spatial patterns}

First, we focus on the pattern formation during the stochastic numerical simulations. After the initial stage of intense predation induced by the random initial conditions, the surviving individuals form two partnerships. This happens because individuals of species $i$, where $i$ is an odd number, join themselves. By their turn, individuals of species $j$, where $j$ is even, are also together. The alliances compete for space, advancing to the adversary territory. The attacks and counter-attacks at the battlefront give rise interfaces of empty sites, bounding the domains.

\begin{figure}[t]
	\centering
	\includegraphics*[width=5cm]{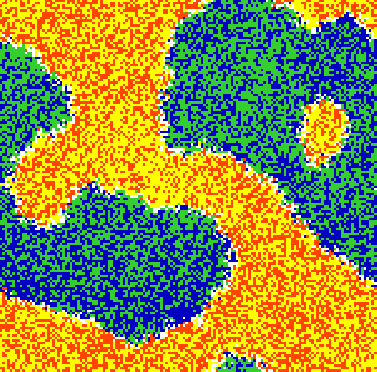}
	\includegraphics*[width=5cm]{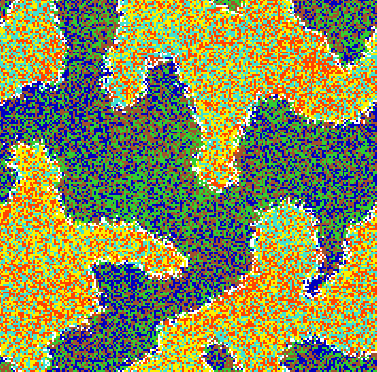}
	\includegraphics*[width=5cm]{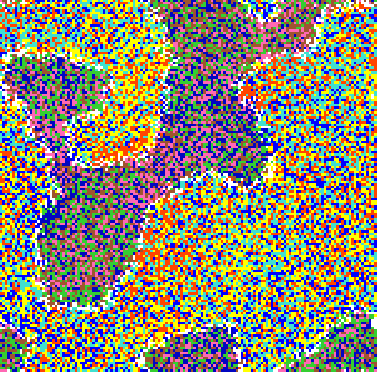}
	\includegraphics*[width=5cm]{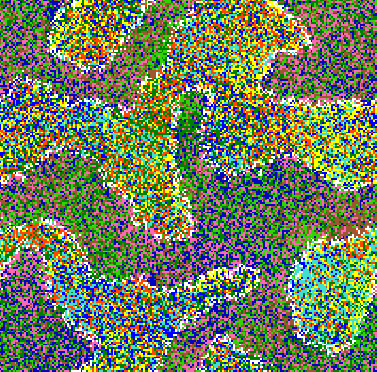}
\caption{Snapshots of stochastic numerical simulations of the $N=4$ (upper left panel), $N=6$ (upper right panel), $N=8$ (lower left panel), and $N=10$ (lower right panel) models. The spatial patterns show spatial domains with partnerships of $N/2$ species, where the individuals are coloured with the respective colour shown in Fig.~\ref{fig1}. The white dots represent the empty sites forming the interface networks, highlighted in Fig.~\ref{fig2a}.}
 \label{fig2}
\end{figure}

Fig.~\ref{fig2} show the spatial patterns resulting from the stochastic numeric simulation for $N=4$ (upper left panel), $N=6$ (upper right panel) , $N=8$ (lower left panel), and $N=10$ (lower right panel) for a $200^2$ network. The snapshots were captured after $200$ generations. The lattice size and the dynamical range were chosen to improve the visualization of the spatial distribution of individuals. Besides, individuals of each species are depicted with the same colors shown in Fig.~\ref{fig1}. For example, for $N=4$ (upper left panel), the alliances $\{1,3\}$ and $\{2,4\}$ arise. In this case, the alliance of red (species $1$) and yellow (species $3$) dots compete with the team formed by green (species $2$) and blue (species $4$) individuals. For the sake of visualization, the snapshots presented in Fig.~\ref{fig2} are only for $N= 4, 6, 8, 10$. \mbox{However}, we have run simulations for large values of $N$ and verified that, in general,  two partnerships are formed (composed of individuals of species $i=1, 3, ..., N-1$ and $i=2, 4, ..., N$). 

\begin{figure}[t]
	\centering
	\includegraphics*[width=5cm]{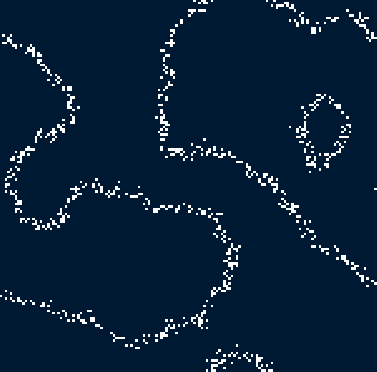}
	\includegraphics*[width=5cm]{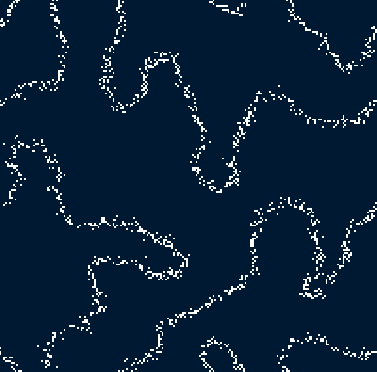}
	\includegraphics*[width=5cm]{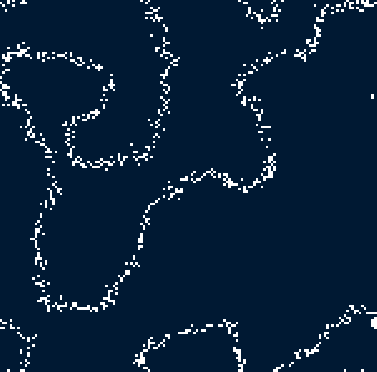}
	\includegraphics*[width=5cm]{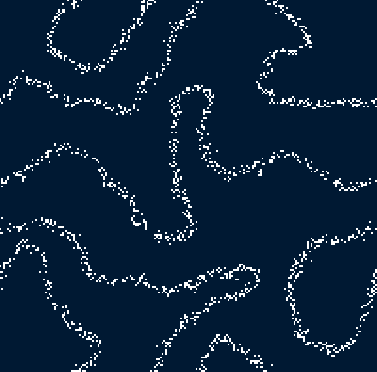}
\caption{Interface networks composed of the empty sites in the snapshots shown in Fig.~\ref{fig2}. The models $N=4$, $N=6$, $N=8$, and $N=10$ are depicted in the upper left panel, upper right panel, lower left panel, and lower right panel, respectively.}
 \label{fig2a}
\end{figure}

Figure \ref{fig2a} highlights the interface network composed of empty sites in \ref{fig2}. The white dots represent the empty sites resulting from the competition activities. They form interfaces of vacancies which are highlighted in the upper left, upper right, lower left, and lower right panels, for $N=4$, $N=6$, $N=8$ and $N=10$, respectively. Whenever an individual is killed at the battlefront, the empty space created can be occupied by offspring of any species. However, this new individual can be caught by an enemy which ensures the stability of the interface. Far away from the interface, there is no empty site since individuals belonging to one alliance do not compete each other.

We point out that although the existence of a model describing the formation of spatial patterns with two competing partnerships \cite{PhysRevE.89.042710}, our model presents two crucial differences. Firstly,  while in that model individuals of various species composing the alliances are homogeneously distributed throughout the respective domains, here the distribution may not be homogeneous. This reflects the way the species interact each other. To be more specific, in that stochastic model \cite{PhysRevE.89.042710}, each species depends on one particular type of teammate to be protected against death threats. This means that individuals of species $i$ only survive in the presence of individuals of species $i-1$, and they are necessary to the persistence of mates of species $i+1$. As a consequence, the individuals that have escaped from being chased in the initial stage of the simulation appear surrounded by their saviors, resulting in homogeneous spatial domains. Conversely, in the model presented here all partners protect each other. As a consequence, individuals of species $i$ may be found surrounded by individuals belonging to any species $j$ of the partnership ($j=i+k$, with even $k$ such that $k=2,4,6,..., N$).  In other words, there is no distinction between the individuals of one alliance. 
This is the reason why clusters of individuals of the same species can be found in the spatial patterns depicted by the upper panels of Fig.~\ref{fig2}. 

Secondly,  in our model, the interfaces do not have internal structures for any number of species $N$. In contrast, in the model in the literature \cite{PhysRevE.89.042710}, the authors claim that for $N \geq 8$, stable dynamical structures are formed at the interfaces due to a peaceful coexistence of species of enemy alliances. These structures are more complex as $N$ increases.
Nonetheless, here individuals of any species compete with everyone from the adversary team. 
Consequently, there is no room for peaceful arrangements at the interfaces. This leads the formation of interfaces without internal structure.   In other words, the model presented in this paper represents a real generalization of interface networks separating two alliances, independent of the number of species. In other words, for the same parameters $m$, $p$, and $r$, the interfaces rising for any $N$ has the same physical properties (thickness and height), as one sees in the four snapshots of Fig.~\ref{fig2}. Later in this paper, we will study the changes in the interface tickness and height for a wide range of $m$, $p$, and $r$.

\section{Mean Field Equations}

In this section, we aim to describe the spatial population dynamics of our stochastic model using a mean field approximation. In this sense, we introduce $N$ real scalar functions $\phi_i(\vec{r}, t)$ (for  $i=1,...,N$), which give
the portion of space around $\vec{r}$ occupied by species $i$ at time $t$. In addition,  $\phi_0(\vec{r}, t)$ gives the percentage of empty space around $\vec{r}$ at the same instant. 

The population dynamics of species $i$ is given by
\bq
\dot{\phi_i} = D\,\nabla^2\,\phi_i + r\,\phi_0\,\phi_i\,-\,p\,\phi_i\,\sum_{\kappa} \phi_{i+\kappa}. \label{egeneral}
\eq
where $i=1,...,N$, and $\kappa$ is an odd number so that $\kappa=1,3, ..., N-1$.  The dot represents a derivative with respect to the physical time. All species have the same diffusion, reproduction and competition parameters, given respectively by $D$, $r$ and $p$, where $D=2\,m$. The number density of empty space is given by $\phi_0=1-\sum_{i}\,\phi_i$. We solved the mean field equations numerically in two spatial dimensions. We assume random initial conditions, where each site is completely occupied by only one species. In other words, we sorted $\phi_i=1$ for $i=1,...,N$, for each grid point. 

\begin{figure}[t]
	\centering
	\includegraphics*[width=5cm]{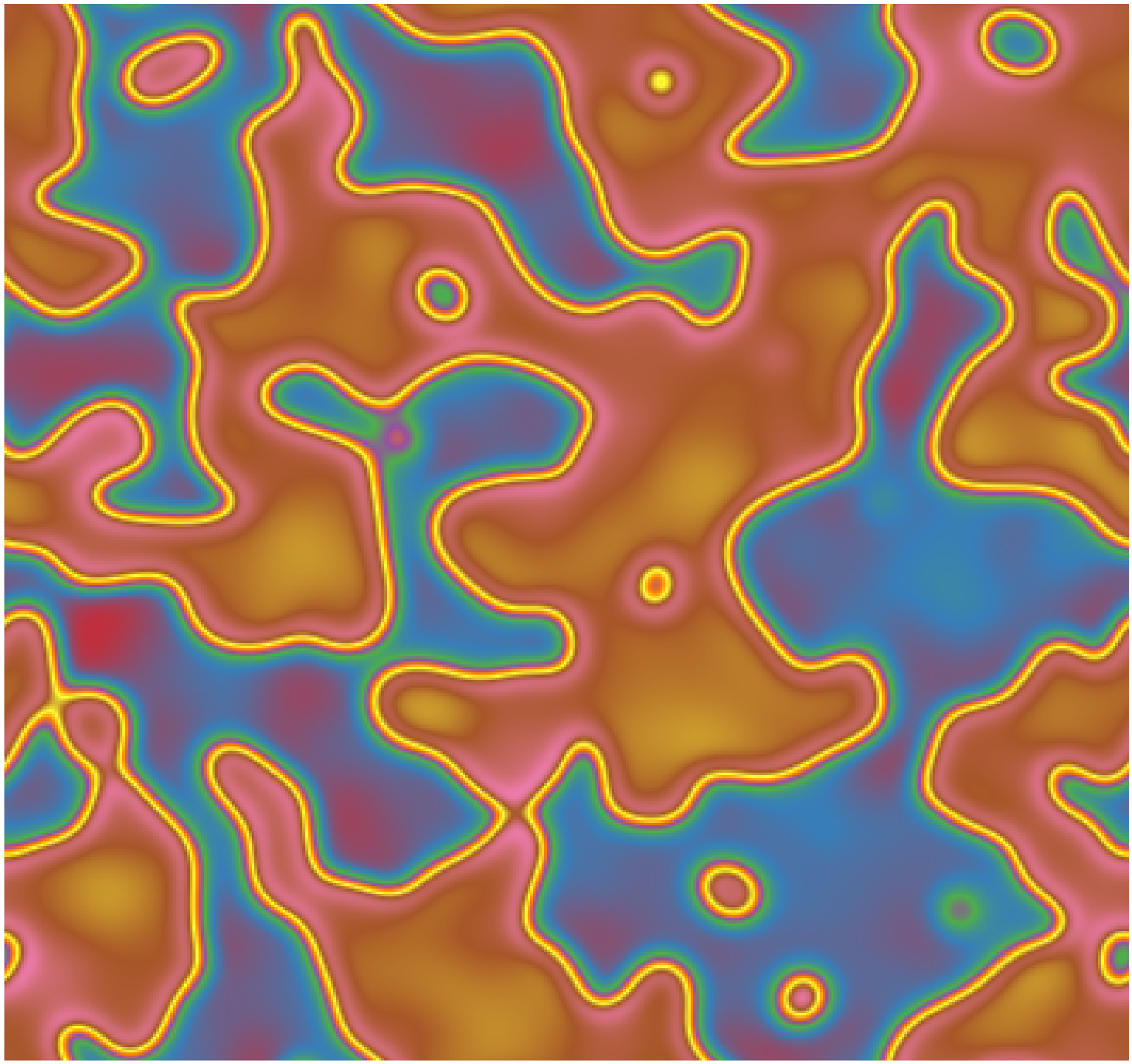}
	\includegraphics*[width=5cm]{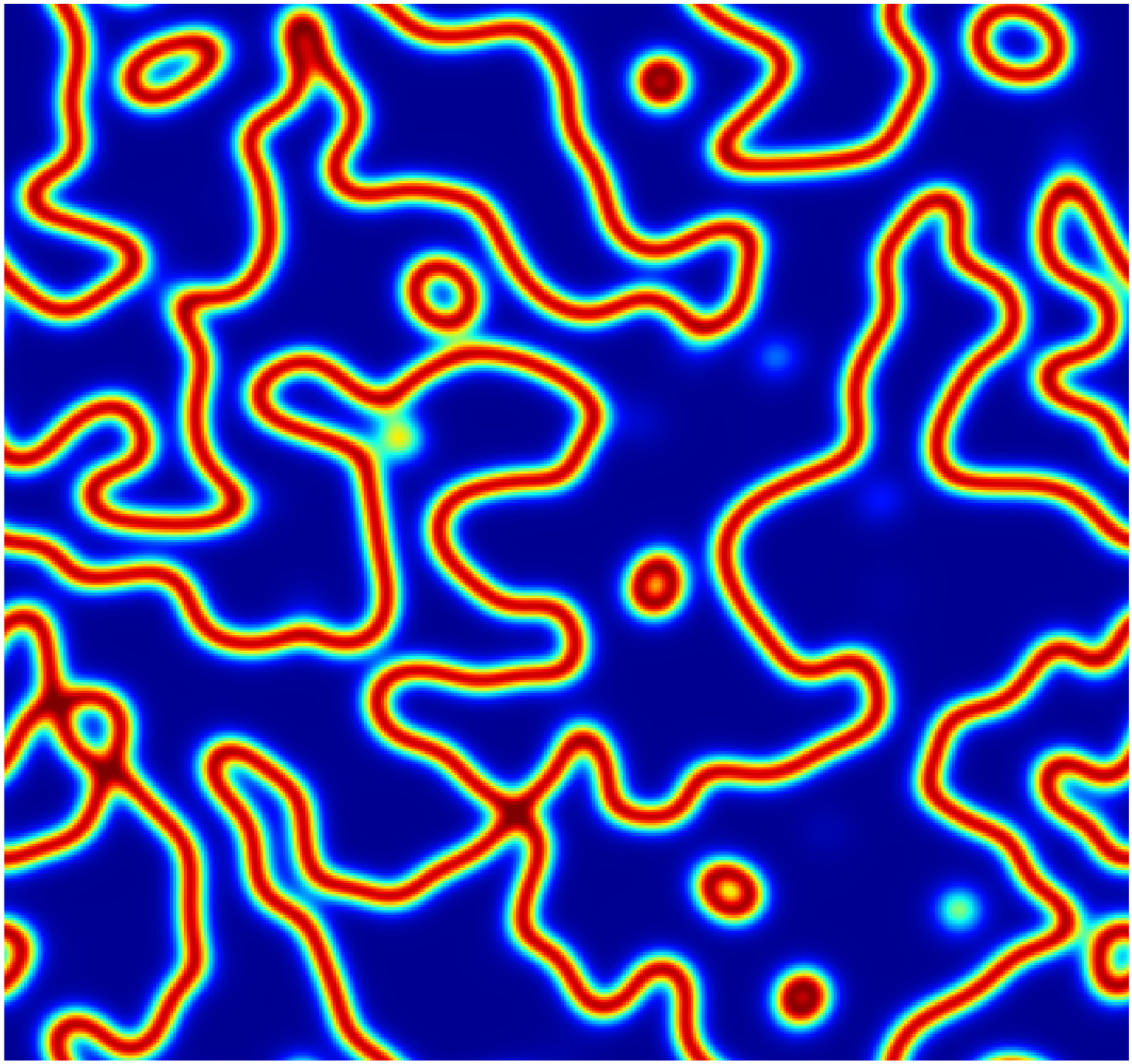}
\caption{Snapshots of $400^2$ two-dimensional implementation of the mean field equations for $N=4$. In the left panel, the colours red and blue represent spatial domains occupied by species $1$ and $3$, while the team of species $2$ and $4$ occupies the pink and brown regions. The right panel depicts the interface network, i.e., the spatial function for $\phi_0$.}
 \label{fig3}
\end{figure}
\begin{figure}
    \center
    \includegraphics[width=10cm]{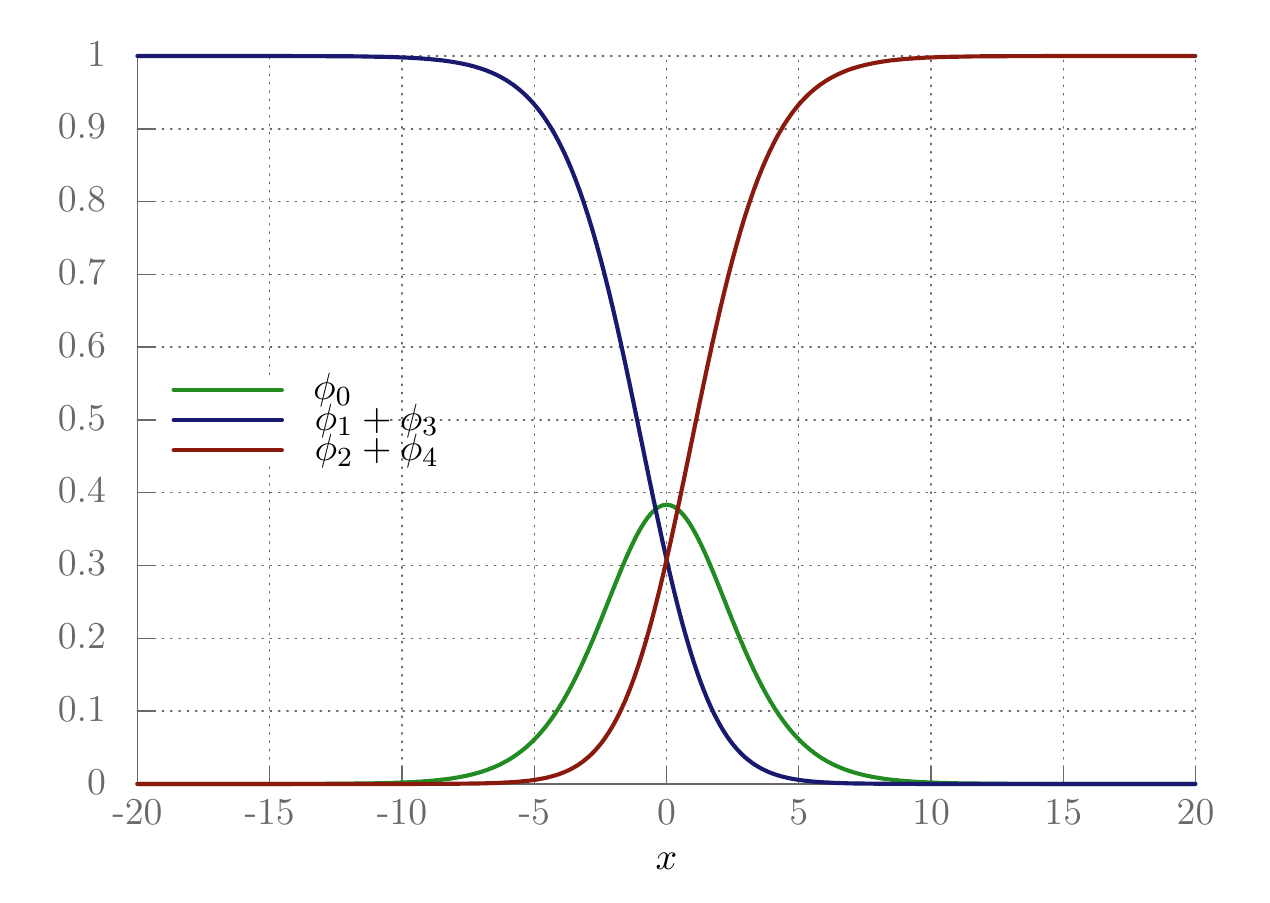}
    \caption{Stationary solutions for the number density of empty space (green line), and partnerships formed by species $\{1,3\}$ (blue line) and species  $\{2,4\}$ (red line).}
     \label{fig4}
\end{figure}

As an example, the left panel of Fig.~\ref{fig3} shows a snapshot of one $400^2$ network for $N=4$, captured at $t=400$.  Here we aim to highlight the spatial distribution of the $\phi_i$ inside the domains of $2$ species that emerge from the initial well-mixed configuration.
The red and blue regions are dominated by the alliance of odd species $\{1, 3\}$, whereas the team $\{2, 4\}$ populates the pink and brown areas. More precisely, one has $\phi_1+\phi_3 =1$ inside the red and blue domains, and $\phi_2+\phi_4 =1$ for the pink and brown regions. This yields the non-homogeneous colors inside the domains. 

The right panel of Fig.~\ref{fig3} shows the interface network of the snapshot in the left panel, where it is highlighted the spatial distribution of number density of empty spaces $\phi_0$. The blue regions are occupied by alliances, such that $\phi_0=0$.
The interfaces tend to straighten so that the total length decreases with time \cite{PhysRevE.86.031119}. As a result, some domains grow while others collapse as a direct consequence of the competition between the partnerships. We notice that either the alliances may coexist at the end of the simulation, or only one invade all territory. In this case, the winner alliance is chosen randomly, since they compete each other with the same parameters. In other words, the surviving species depend on the random initial network configuration.

\subsection{The interface profile}

To see how $\phi_0$ behaves as the interface, we have solved the mean field equations for one spatial dimension. We observe the number density of empty space ($\phi_0$) at the interface does not depend on the number of species $N$. For example, for $N=4$, the one dimensional implementation of the mean field equations started from initial conditions where $\phi_1+\phi_3=1$  for $x<0$ and $\phi_2+\phi_4=1$ for $x>0$. After a few number of time steps, competition interactions give rise a stable interface at $x=0$,  as it is shown by the green line of Fig.~\ref{fig4}.
This static one-dimensional solution for the number density of empty space is defined as the interface profile. This implies that the interface height $H$ is given by $\phi_0$ at $x=0$. Moreover, at the interface center one has $\phi_1+\phi_3=\phi_2+\phi_4=(1-H)/2$. In addition, $\phi_1+\phi_3=1$ and  $\phi_2+\phi_4=1$ asymptotically for $x \to -\infty$ and $x \to \infty$, respectively.  This reveals the topological property of the interface profile, as it is depicted by the blue and red lines of Fig.~\ref{fig4}, respectively.

\begin{figure}[t]
    \center
    \includegraphics[width=10cm]{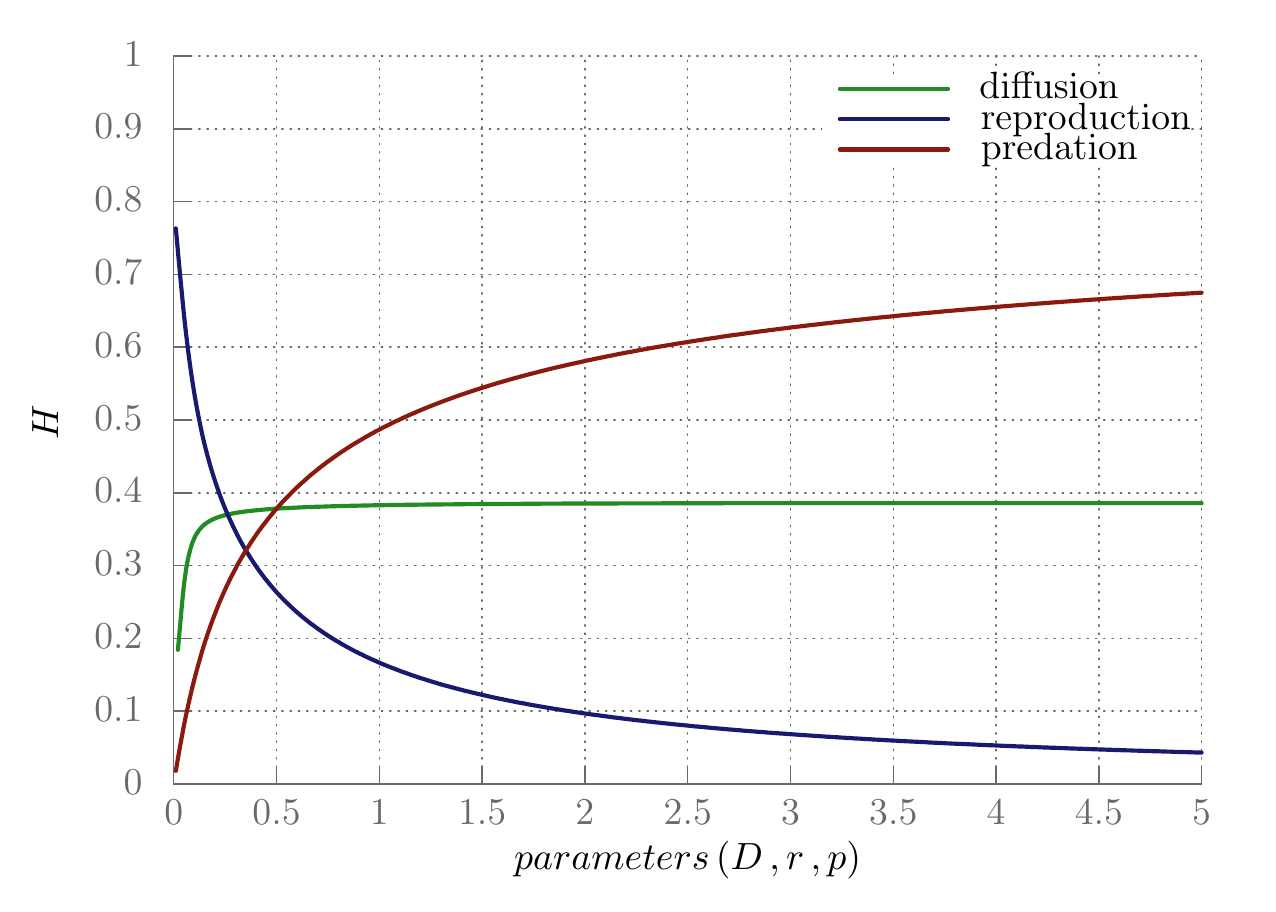}
    \caption{Variation of the interface height by changing one parameter at a time. The red line shows $H$ for varying $p$, and fixed $D$ and $r$ ($D =0.50$ and $r=0.25$). Similarly, the green and blue lines show the results for varying $D$ ($r =0.25$ and $p=0.50$) and $r$ ($D =0.50$ and $p=0.50$), respectively. The inset panel shows that $H$ changes slightly as $D$ increases.}
    \label{fig5}
\end{figure}

Furthermore, we verified how the interface height depends on the model parameters. To this purpose, we measure numerically $H$ for stable interfaces, by varying one of the parameters $(D, r, p)$ while two of them remain constant in their respective values $(0.50, 0.25, 0.50)$. The results shown in Fig.~\ref{fig5} suggest that the larger $p$, the more intense is the predation activity, which leads to a higher interface. Conversely, the more the individuals reproduce (larger $r$), the lower the interface is.

\subsection{The dynamics of the interface networks}

We focus now on the macroscopic evolution of the interface networks. Our goal is to find out how the interface networks evolve for different values of $N$. The characteristic length $L$ of the network is defined as the ratio between the area of the lattice $A$ (the total number of grid points), and the total length of interfaces $L_T$, i.e., $L = \frac{A}{L_T}$. Given that the interface thickness is fixed throughout the lattice, the number density of empty spaces per cross section area does not vary. As a consequence, $L_T$ is proportional to the total number of empty space. Therefore,  $L$ is inversely proportional to the sum of $\phi_0$ for the entire network \cite{PhysRevE.89.042710}.

\begin{figure}[t]
    \center
    \includegraphics[width=10cm]{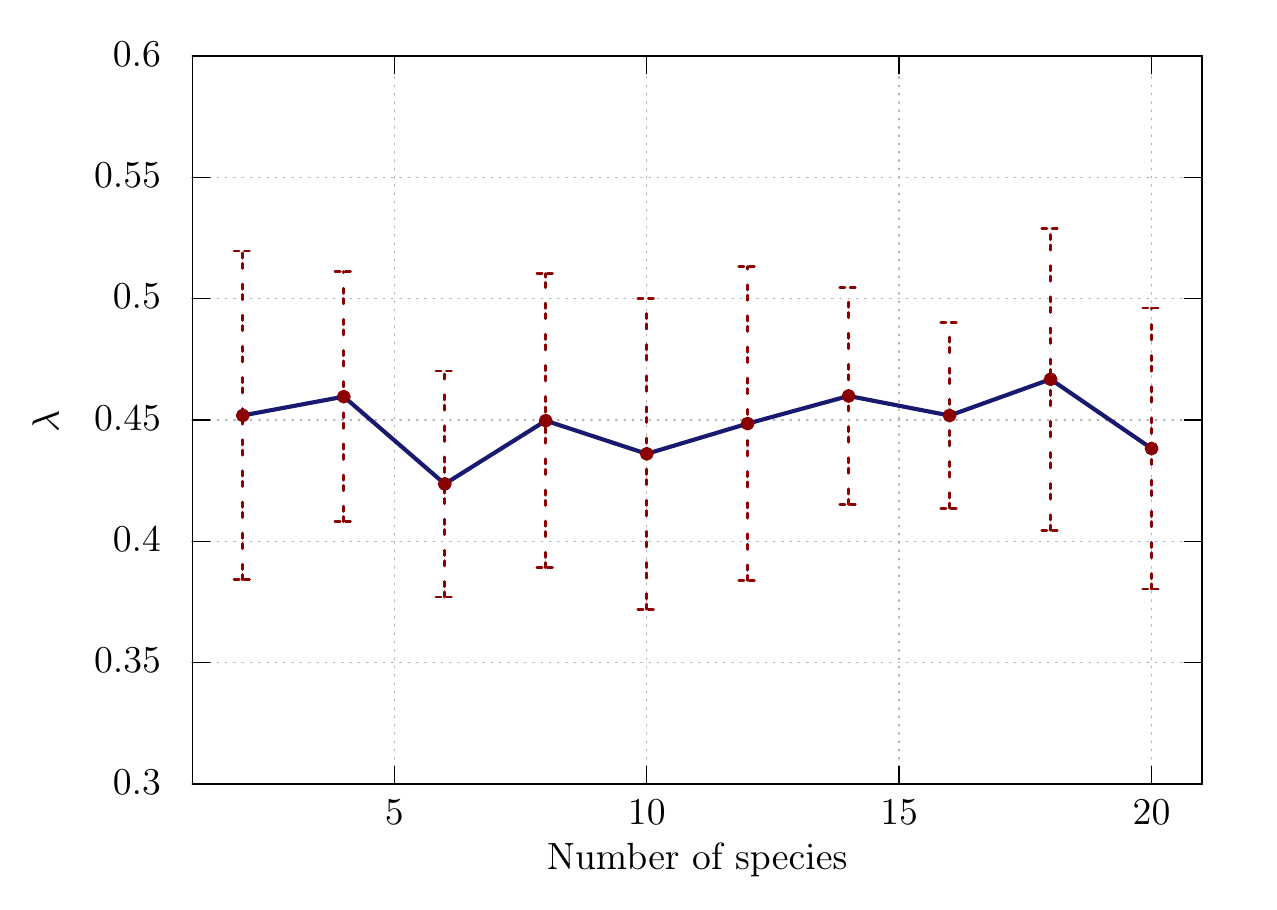}
    \caption{Scaling exponent for two-dimensional numerical implementations of the mean field for $2 \leq N\leq 20$. The error bars indicate the standard deviation $1\sigma$ of the average value of $\lambda$ taken from $25$ realizations.}
     \label{fig6}
\end{figure}

We computed the average evolution of $L$ with time by running series of $25$ two-dimensional mean-field simulations with distinct initial conditions for $2 \leq N\leq 20$. 
We verified that the scaling law $L \propto t^{\lambda}$ describes well the late time evolution of the interface networks for any $N$, which is similar to curvature-driven network evolution of nonlinear systems \cite{PhysRevE.86.031119}.

The results are shown in Fig.~\ref{fig6}, where the error bars indicate the standard deviation $1\sigma$ of the average value of $\lambda$.  One sees that  $\lambda \approx 0.5$, which indicates that the macroscopic evolution of the interface networks is independent of the number of species composing the alliances. The deviation of $\lambda$ to $0.5$ is also present in previous studies of dynamics of interface networks, where the scaling exponent is approximately the same presented here \cite{PhysRevE.86.031119}. This shows that our mean field model describes well the dynamics of the stochastic system. However, the reason this significant deviation occurs will be investigated in further work.
\section{Soliton Topological Model}
In this section we objective to write a soliton topological model to describe the appearance of the interface networks as a result of the spontaneous breaking of a discrete symmetry \cite{rajaraman1982solitons}. The numerical results presented in the previous sections show that an interface is a boundary between spatial regions populated by two adversary alliances. 
Therefore, we introduce a real scalar field defined as 
\be
\Phi=\sum_{j}\,\left(\phi_{j+1}-\phi_{j}\right)
\ee
where $j$ is an odd number, such that $j=1,3,..., N-1$. This allows the replacement of the mean field equations (Eq. \ref{egeneral}) by the field equation
\be
\dot{\Phi} = D\,\nabla^2\,\Phi - r\,\phi_0\Phi.
\label{allen}
\ee
In this approach, the alliances arise for $\Phi(x \to \pm \infty) = \pm 1$. For example, for $N=4$ one has $\phi_2+\phi_4=1$ for $\Phi=-1$,  while  $\Phi=1$ is assumed when $\phi_1+\phi_3=1$.
We highlight that Eq. \ref{allen} can be interpreted as the Allen-Cahn equation, which is a reaction-diffusion equation widely used to investigate phase separation in alloy systems  \cite{allen}. In this case, the phase transition is determined by a potential $V(\Phi)$ so that $d\,V/d\,\Phi= - r\,\phi_0\Phi$.
\begin{figure}[t]
   \center
    \includegraphics[width=10cm]{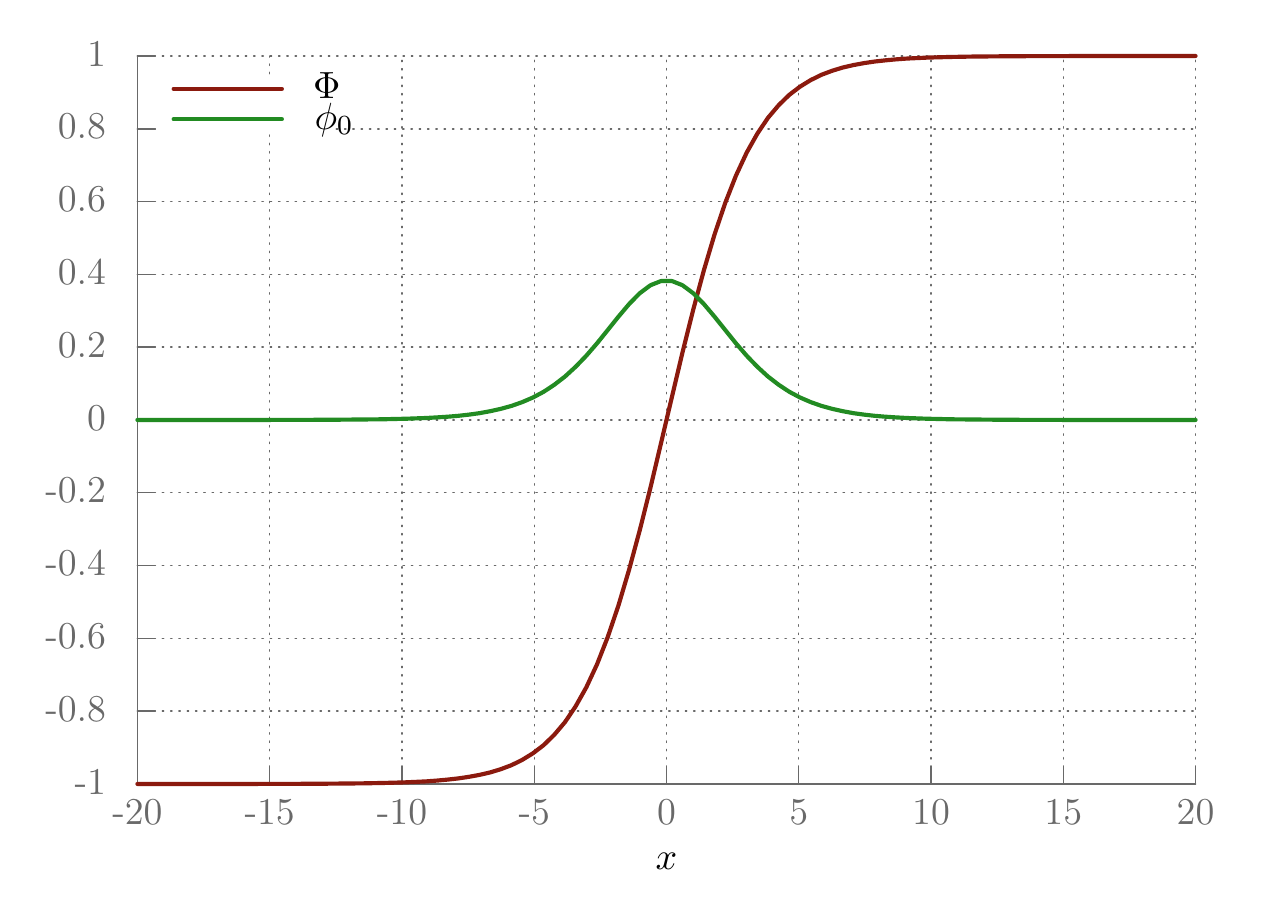}
    \caption{The red line shows the positive soliton solution $\Phi(x)$ for the static, one-dimensional version of the field equation. The stationary interface profile $\phi_0$ is represented by the green line.}
     \label{fig7}
\end{figure}

Let us now focus on the one-dimensional static solution of Eq. \ref{allen} to write an analytic function for the stationary interface profile. Taking into account the existence of two asymptotic partnerships far away from the interface center, we assume the potential with discrete symmetry given by
\be
V(\Phi)= \eta\,\left(1-\Phi^2\right)^2,
\label{potential}
\ee
where $\eta$ is a function of $D$, $r$ and $p$.
The interface is the barrier separating the alliances represented by the potential minima $\Phi=-1$ (even species) and $\Phi=1$ (odd species).
Therefore, the static soliton solution for Eq. \ref{allen} is
\be
\Phi(x)=\pm \tanh \left(\sqrt{\frac{2\,\eta}{D}}\,x \right), \label{solitonic}
\ee   
which asymptotically connect the potential minima. 
The positive solution represents $(\phi_1+...+\phi_{N-1})$ going from $1$ to $0$ through the interface, while the negative solution gives the opposite asymptotic behavior.  Accordingly, the analytic expression for the interface profile $\phi_0$ is
\be
\phi_0(x)\,=\,H-H\,\tanh^2\left(\sqrt{\frac{2\,\eta}{D}}\,x \right), \label{eq.phi0-x}
\ee
whose thickness is given by
$\delta \simeq \pi/\sqrt{2\,\eta}$  \cite{1989ApJ...347..590P}.

Figure~\ref{fig7} shows the positive solution $\Phi$ and the interface profile for the set of parameters assumed in the previous sections. Note that $\eta\,=\,H\,r/4$, where $H$ is the interface height computed in the one-dimensional implementation of the mean field equations. The parameters $\eta$ and $D$  determine how fast the solutions reach the potential minima. 

\begin{figure}[t]
	\centering
	\includegraphics*[width=5cm]{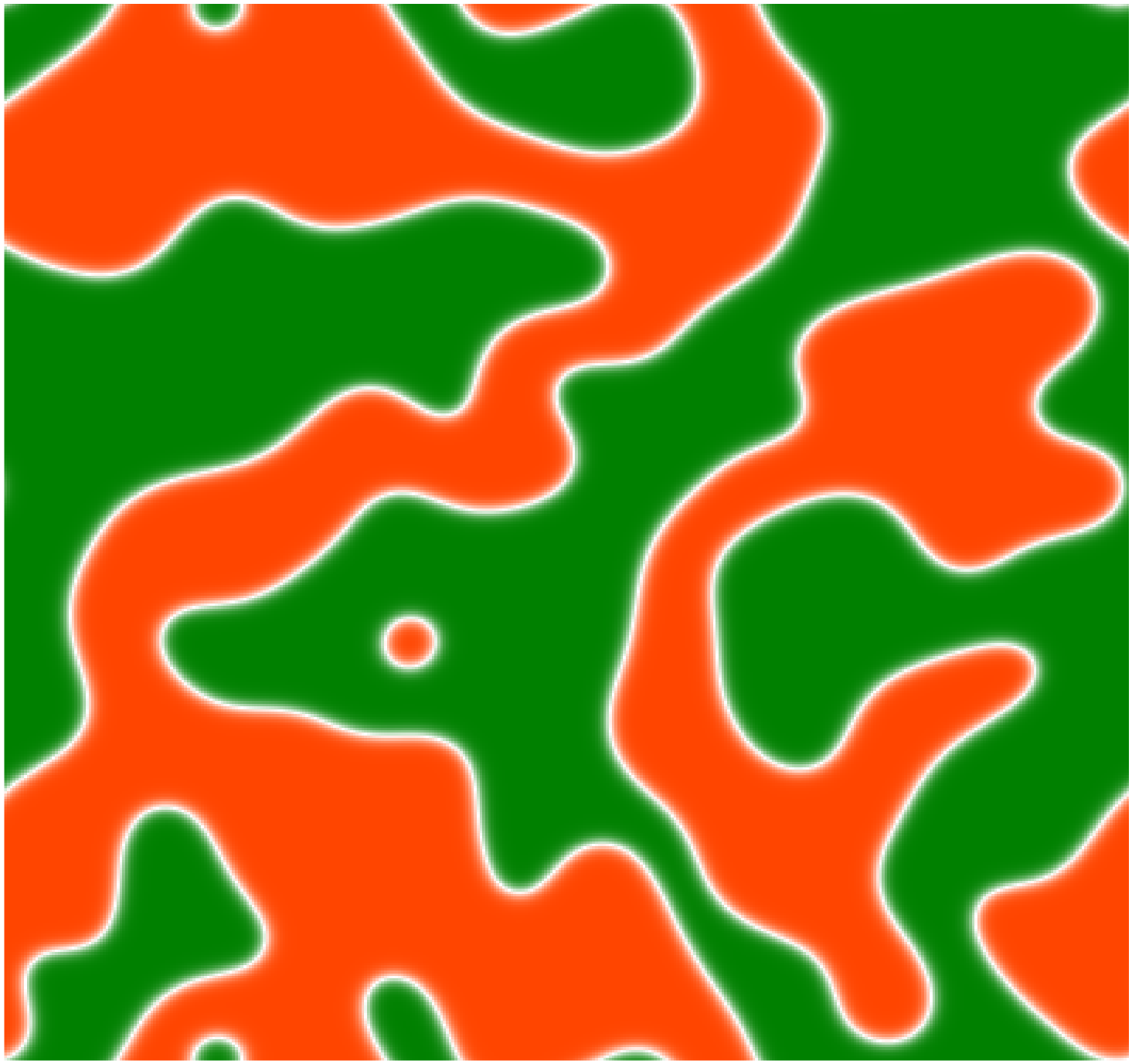}
	\includegraphics*[width=5cm]{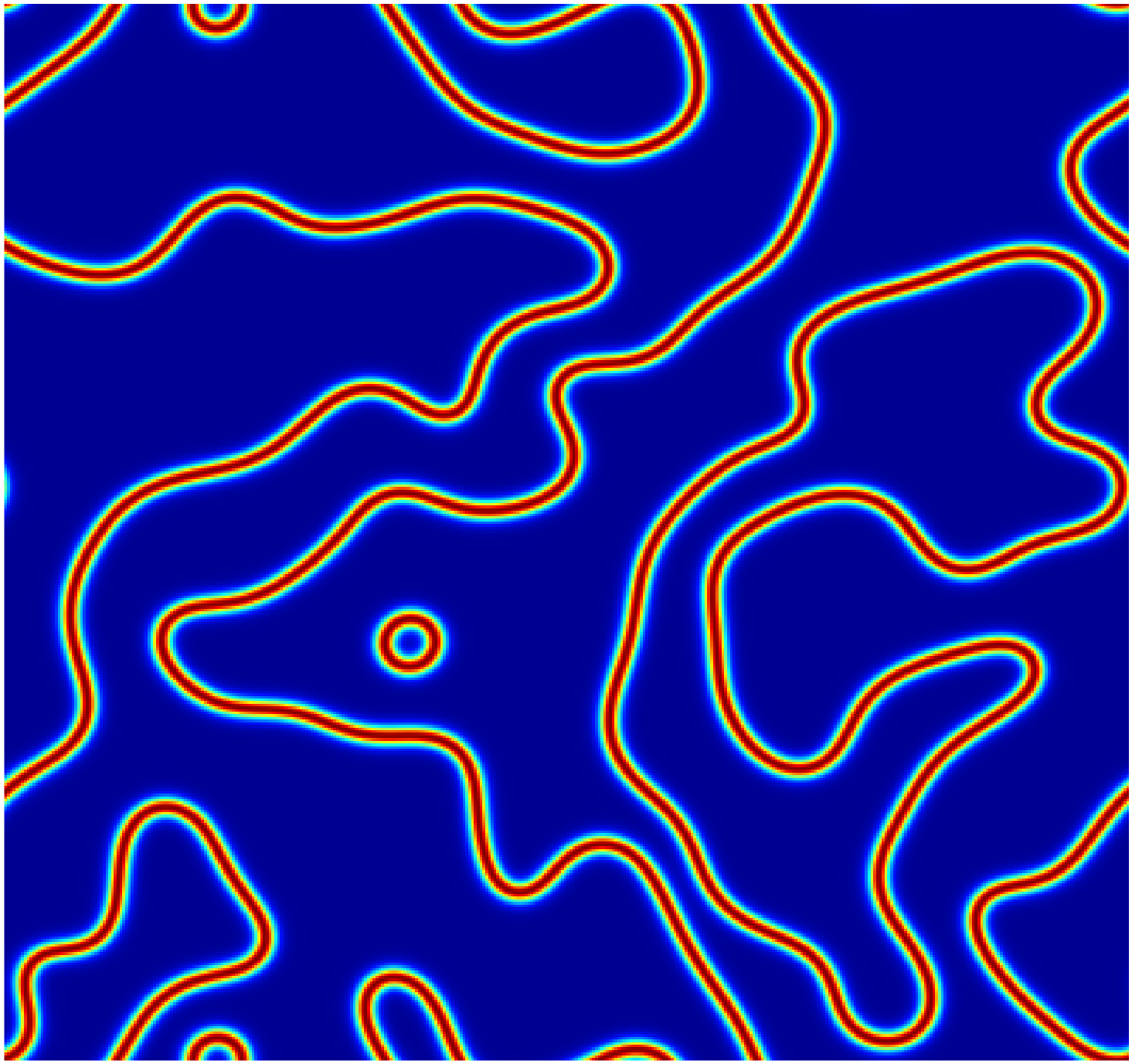}
\caption{Snapshots of the implementation of the field equation in a $400^2$ two-dimensional network. In the left panel, the colors green and orange represent
spatial domains occupied by even and odd species, respectively. The red lines in the right panel show the interfaces.}
  \label{fig8}
\end{figure}

We solve Eq. \ref{allen} numerically in two spatial dimensions, assuming the potential in Eq. \ref{potential}. The grid points are given random initial values for the $\Phi$ between $-1$ and $1$. 
The well-mixed spatial initial configuration disappears after a few number of step times when spatial domains of alliances arise  \cite{video1,video2}.  In other words, spatial domains with $\Phi=-1$ or $\Phi=1$ are created, as it is shown in the left panel of Fig. \ref{fig8}. The snapshot was taken from a $400^2$ grid, at $t=400$. The domains are bounded by the interfaces shown in the right panel of Fig. \ref{fig8}.

Finally, by carrying out $25$ numerical implementations of the field equation, we found out that the interface network change in time according to scaling law $L \propto t^{\lambda}$, with $\lambda=0.444 \pm 0.07$. This agrees with the results obtained by the numerical implementation of the mean field equations presented in the previous section.

\section{Conclusion and Discussion}

Performing numerical simulations of a subclass of the more general May-Leonard stochastic model, we found that the spatial patterns show the appearance of spatial regions populated by two competing partnerships.  Each partnership member is responsible for keeping partners safe and contributing to the territorial advance of the alliance. In other words, there is no distinction between individuals of one partnership. As a result, interface networks without internal structures arise irrespective of the number of species present.

We solved the mean field equations associated to the stochastic model numerically, in one and two spatial dimensions. The results describe well the spatial patterns provided by the stochastic simulations. The interface profile shows the topological aspects of the system. This means that far away from the interface core, space is occupied for enemy alliances. Based on this, we presented a theoretical approach, where the arising of spatial patterns result from the spontaneous breaking of a discrete symmetry. On the hand, the analytic solution of the field equation allows to the description of the topological properties of the stationary interface profiles. The spatial patterns resulting from the two-dimensional implementation of this approach strongly agree with the numerical results provided by the mean field model and the stochastic numerical simulations.
The dynamics of the interface was also computed using the soliton topological model, showing that characteristic length of the interface networks evolves according to the same scaling law of curvature driven system in various nonlinear systems.

Furthermore, this formalism can be generalized to models with competing species whose interactions lead to the formation of string networks with and \mbox{without} junctions \cite{Avelino2014393,Menezes2017}. The same topological issue is present in this case, except for the spatial symmetry engendered by the spatial distribution of the species around the defect cores. In this case, a potential with continuous symmetry is needed, so that the field equation can describe the formation and evolution of the string networks. It has been shown numerically that the string profile is a function of the parameters $D$, $r$ and $p$, which is in complete agreement with the analytical results presented here \cite{Menezes2017}.

Finally, we point out that the process of phase transition resulting from a spontaneous symmetry \mbox{breaking} is largely studied in defect networks in Cosmology and \mbox{Condensed} Matter \cite{vilenkin}. The application of the same principles for explaining spatial patterns in \mbox{Biology} constitutes a tool for theoretical researchers, bringing novelty and inspiration to use the physical knowledge to understand old problems regarding the coexistence among species.

\section*{Acknowledgements}
This study was supported by CAPES, CNPq, FAPERN, and the Netherlands Organisation for Scientific Research (NWO) for financial and computational support. JM acknowledges support from NWO Visitor's Travel Grant 040.11.643.

\end{document}